\documentclass[traditabstract]{aa}

\usepackage{amsfonts}
\usepackage{amsmath}
\usepackage{amssymb}
\usepackage{pdflscape}
\usepackage{rotating}
\usepackage{natbib}
\usepackage{float}
\usepackage{adjustbox}
\usepackage{txfonts}
\usepackage{longtable}
\usepackage[small,bf]{caption}
\usepackage[small,bf]{subfigure}
\usepackage{boldline,multirow}
\usepackage{xspace}
\usepackage{graphicx}
\usepackage{epstopdf}
\usepackage[pdftitle={3c120}, colorlinks=true, citecolor=blue, linkcolor=blue, urlcolor=blue]{hyperref}

\newcommand\simlt{\lower.5ex\hbox{$\; \buildrel < \over \sim \;$}}

\newcommand{\gray}{$\gamma$-ray\xspace}
\newcommand{\grays}{$\gamma$-rays\xspace}

\DeclareUnicodeCharacter{B0}{\textdegree}

\begin{document}
\title{Comparing 3C 120 jet emission at small and large scales}
\titlerunning{X- and Gamma-ray emission from the jet of 3C 120}

\author{D. Zargaryan
\and S. Gasparyan
\and V. Baghmanyan
\and N. Sahakyan
}

 \institute{ICRANet, Piazza della Repubblica 10, I-65122 Pescara, Italy
 \and ICRANet-Armenia, Marshall Baghramian Avenue 24a, Yerevan 0019, Republic of Armenia }

  \abstract
   {Important information on the evolution of the jet can be obtained by comparing the physical state of the plasma at its propagation through the broad-line region (where the jet is most likely formed) into the intergalactic medium, where it starts to significantly decelerate.}
   {We compare the constraints on the physical parameters in the innermost ($\leq$ pc) and outer ($\geq$ kpc) regions of the 3C 120 jet by means of a detailed multiwavelength analysis and theoretical modeling of their broadband spectra.}
   {The data collected by Fermi LAT (\gray band), Swift (X-ray and ultraviolet bands) and Chandra (X-ray band) are analyzed together and the spectral energy distributions are modeled using a leptonic synchrotron and inverse Compton model, taking into account the seed photons originating inside and outside of the jet. The model parameters are estimated using the Markov Chain Monte Carlo method.}
   {The \gray flux from the inner jet of 3C 120 was characterized by rapid variation from MJD 56900 to MJD 57300. Two strong flares were observed on April 24, 2015 when, within 19.0 minutes and 3.15 hours the flux was as high as $(7.46\pm1.56)\times10^{-6}\;{\rm photon\:cm^{-2}\:s^{-1}}$ and $(4.71\pm0.92)\times10^{-6}\;{\rm photon\:cm^{-2}\:s^{-1}}$ respectively, with $\geq10\sigma$. During these flares the apparent isotropic \gray luminosity was $L_{\gamma}\simeq(1.20-1.66)\times10^{46}\:{\rm erg\:s^{-1}}$ which is not common for radio galaxies. The broadband emission in the quiet and flaring states can be described as synchrotron self-Compton emission while inverse Compton scattering of dusty torus photons cannot be excluded for the flaring states. The X-ray emission from the knots can be well reproduced by inverse Compton scattering of cosmic microwave background photons only if the jet is highly relativistic (since even when $\delta=10$ still $U_{\rm e}/U_{\rm B}\geq80$). These extreme requirements can be somewhat softened assuming the X-rays are from the synchrotron emission of a second population of very-high-energy electrons.} 
 {We found that the jet power estimated at two scales is consistent, suggesting that the jet does not suffer severe dissipation, it simply becomes radiatively inefficient.}

\maketitle

\section{Introduction}
The recent observations in the High Energy (HE; $\geq100\:{\rm MeV}$) \gray band show that the extragalactic \gray sky is dominated by the emission of Active Galactic Nuclei (AGNs) of different types. Dominant in these are blazars - an extreme class of AGNs which have jets that are forming a small angle with respect to the line of sight \citep{urry}. Blazars are very bright and luminous sources known to emit electromagnetic radiation in almost all frequencies that are currently being observed, ranging from radio to Very High Energy (VHE; $>$ 100 GeV) \gray bands. Their broadband spectrum is mainly dominated by non-thermal emission produced in a relativistic jet pointing toward the observer. Other important class of \gray emitting AGNs observed by Fermi Large Area Telescopes (Fermi LAT) are radio galaxies with relativistic jets at systematically larger angles \citep{abdomisaligned,katak}. Due to larger jet inclination angle as compared with blazars, the jet emission is not significantly Doppler boosted, making it less prevalent over such components as the radiation from mildly relativistic outflows or emission from extended structures. This opened a new window to have an insight into the particle acceleration and emission processes in different components of AGNs.\\
The radio galaxy 3C 120, at a distance of $\approx144.9$ Mpc, is an active and powerful emitter in all the observed wavebands. In the radio band, its characteristics are closer to the Fanaroff-Riley class I radio sources \citep{fanaroff74} with a powerful one-sided radio jet from sub-$pc$ to 100 $kpc$ scales \citep{walker}. The one-sided parsec-scale jet has been studied by long baseline interferometry and superluminal motion has been observed with apparent speed up to 4-6 $c$ \citep{homan01,gomes98,gomes99}. Recently, using X-ray and radio observations, \citet{mars02} found that the dips in the X-ray emission are followed by ejections of bright superluminal knots in the radio jet which clearly establishes an accretion-disk-jet connection. The kpc-scale jet of 3C 120 has a complex structure with several knots, k4, k7, s2, s3, and k25 (see Fig. 2 of \citet{hariss04}, where on the 1.6 GHz radio contours the section of the jet with knot labels is shown), detected in the radio, optical, and X-ray bands \citep{HARIS99,hariss04}. The knots are labeled by their distance from the core in arcseconds (e.g., k4, k7) and the smooth sections of the jet detected in the optical band \citep{hjorth} are labeled as s2 and s3. These knots appeared to have interesting morphology and spectra, more tricky among which is the X-ray emission from the knot k25: it has a very weak radio flux but it is bright in the X-ray band. It is a real challenge for one-zone synchrotron emission scenario to interpret the emission from k25 in case of which large deviation from the minimal energy condition is required. It has been suggested that X-rays might be produced through the synchrotron radiation of an electron population distinct from that responsible for the radio emission \citep{hariss04}. Alternative theories such as proton synchrotron emission \citep{ahar2002} or inverse-Compton scattering of CMB photons \citep{zhang} have also been proposed. However, it is to date not clear which is the exact mechanism responsible for the X-ray emission.\\
The core of 3C 120’s jet itself has interesting and peculiar features. It is very bright in the X-ray band with a flux of $\approx5\times10^{-11}\:\mathrm{erg\:cm^{-2}s^{-1}}$ at 2-10 keV, variable on time scales from days to months \citep{halpern}. The \grays from 3C 120 had been already detected by Fermi LAT during first 15 months’ scan of the whole sky \citep{abdomisaligned} which was then confirmed by the data accumulated for two years \citep{katak}. Also a long-term (several months) variability had been found using the five-year Fermi LAT data \citep{sahak3c120} with short periods (days and hours) of brightening \citep{tanaka3c120,sikora16}. Inverse Compton scattering of synchrotron photons seems to be the mechanism responsible for the \gray emission from 3C 120 \citep{tanaka3c120,sahak3c120} while the flares and the fast \gray variability are explained within more complex structured jet scenarios \citep{sikora16}.\\
Combining of the data derived at the sub-$pc$ and kilo-parsec regions of the same jet could greatly help to understand the features of powerful extragalactic jets, e.g., shed some light on the evolution and propagation of the jets from the central engine to the outer regions, where the jet is starting to significantly decelerate. This approach can be fruitfully applied to the sources showing a large-scale jet long enough to be resolved by Chandra. Unfortunately, the best-studied blazars do not tend to have well-studied large-scale jets, precisely because the blazars are most closely aligned with the line of sight, reducing the projected angular dimension of the large-scale jet. Thus, only a few jets can be studied on both scales. The prominent features of 3C 120, e.g., the strong jet well resolved in both small (pc) and large (kpc) scales makes this object an ideal target for investigation of the processes occurring in the powerful jet along its propagation.\\
The low statistics in the \gray band did not allow to study the flux changes on sub-month scales (the light curves contain many upper limits). The recent update of the Fermi LAT event-level analysis from {\it PASS7} to {\it PASS8} has significantly improved the event reconstruction and classification which increased the sensitivity and improved the angular resolution. Combining this with the data accumulated for a longer period (8 years) would significantly improve the statistics allowing to perform a detailed study of the \gray flux evolution in time. Also, the analysis of the Swift data will allow to explore the emission from the core region with the help of contemporaneous Spectral Energy Distributions (SEDs). Moreover, due to several Chandra observations of the large-scale jet of 3C 120 in 2001-2016, the overall exposure is large enough to perform a spectral analysis of the X-ray data. This motivated us to have a new look on the emission from the jet of 3C 120 in small and large scales using the most recent data available.\\
The paper is structured as follows. The analyses of Fermi LAT and Swift XRT/UVOT data are presented in Section \ref{sec2}. The analysis of Chandra data is described in Section \ref{sec3}. The modeling of the emission from the core and knots is presented in Section \ref{sec4}. The results are discussed in Section \ref{sec5} and summarized in Section \ref{sec6}.
\begin{figure*}[!ht]
   \centering
  \includegraphics[width=0.96 \textwidth]{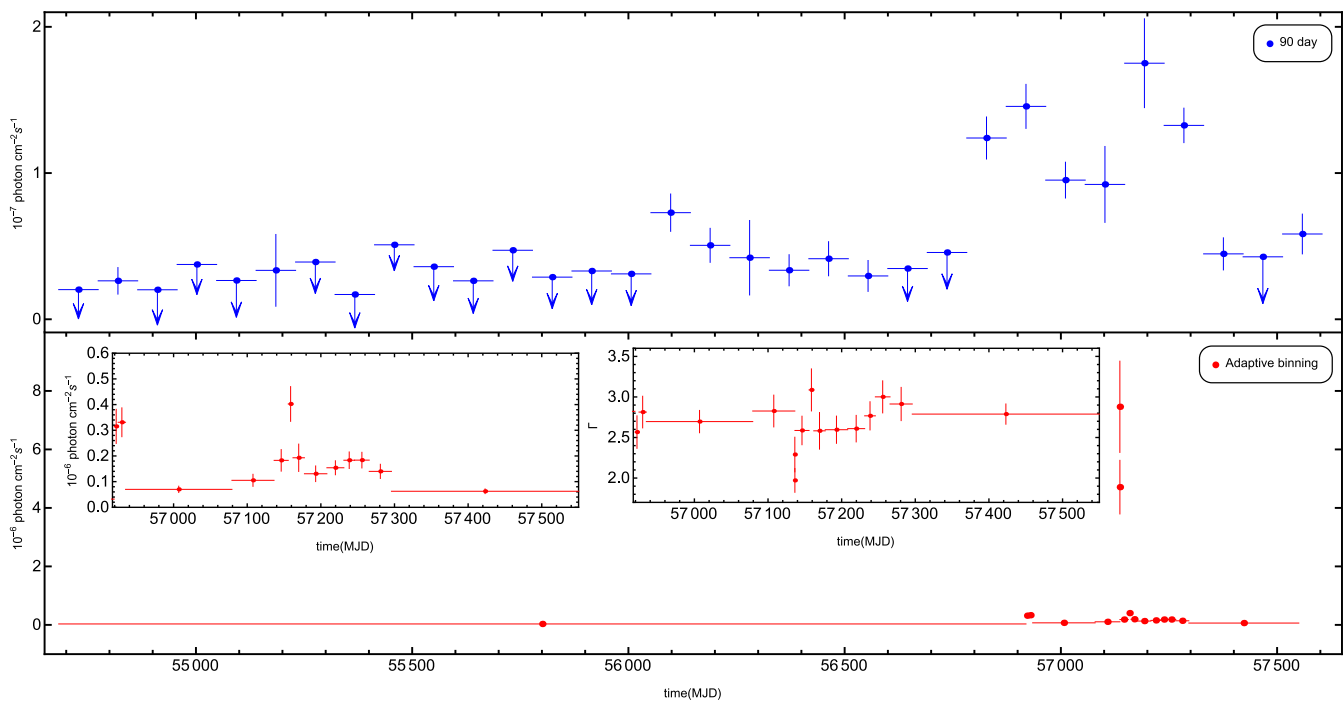}
   \caption{The \gray light curve of 3C 120 from August 4, 2008, to August 4, 2016. (a) The bin intervals correspond to 90 days. (b) The light curve obtained by adaptive binning method assuming 20 \% of uncertainty. The change of photon index is shown in the insert.}
    \label{fg2}
\end{figure*}
\section{Observations and data analysis of central region}\label{sec2}
\subsection{Fermi LAT data extraction}
On board the Fermi satellite, LAT is a pair-conversion telescope designed to detect HE \grays in the energy range 20 MeV - 300 GeV. Operating since August 4, 2008, it is always in the survey mode by default, scanning the entire sky every $\sim$3 hours, thereby providing continuous monitoring of \gray sources. Details about the Fermi LAT can be found in \citet{atwood}.\\
In the current paper we analyze the data accumulated for 8 years, from the beginning of the Fermi LAT mission up to August 8, 2016 (MET 239557417-460339204). Fermi LAT Science Tool version v10r0p5 is used with the instrument response function P8R2\_SOURCE\_V6. The recently updated {\it PASS8} version of the data in the energy range between 100 MeV - 300 GeV is analyzed. The entire data set is filtered with {\it gtselect} and {\it gtmktime} tools and only the events with a high probability of being photons {\it evclass=128, evtype=3} have been considered. The zenith angle cutoff $>90^{\circ}$ is made to exclude atmospheric \grays from the Earth limb that can be a significant source of background.\\
The photons from a circular Region of Interest (ROI) centered on VLBI radio positions of 3C 120 (RA,dec) = (68.296, 5.354) are used in the analysis. The photons are binned within $14.1^{\circ}\times14.1^{\circ}$ square regions with {\it gtbin} tool, with a stereographic projection into $0.1^{\circ}\times0.1^{\circ}$ pixels. In order to account for the emission from other sources within ROI, the model file is generated using the Fermi-LAT third source catalog \citep{acero15} and the sources within $10^{\circ}+5^{\circ}$ from the position of 3C 120 are included in the model file. Since 3C 120 is not included in the catalog, a point-like source in the known location of 3C 120 was added to the model file. The Galactic background component is modeled using the LAT standard diffuse background model {\it gll\_ iem \_ v05\_ rev1} and {\it iso\_source\_v05} – for the isotropic \gray background. The normalization of background models as well as the fluxes and spectral indices of the sources within $10^{\circ}$ are left as free parameters during the analysis.
\subsection{Temporal variability}
In order to have SEDs with contemporaneous data for broadband modeling we created \gray light curves with different time binning. The \gray light curve is calculated with the unbinned likelihood analysis method implemented in the {\it gtlike} tool. $(0.1-300)$ GeV photons are used in the analysis with the appropriate quality cuts applied in the data selection. The photon indices of all background sources are fixed to the best guess values obtained in full time analysis in order to reduce the uncertainties in the flux estimations. The power-law index of 3C 120 is first considered as a free parameter and then as a fixed one. Since no variability is expected for the background diffuse emission, the normalization of both background components is also fixed to the values obtained for the whole time period.\\
The light curve obtained for 90-day binning is shown in the upper panel of Fig. \ref{fg2}. Before $\approx$ MJD 56000 (March 14, 2012), the source is mostly undetectable by Fermi LAT: only in two of the total 15 cases, the source detection significance exceeded the required threshold of $4\sigma$. Then the source flux was high enough to be detected by Fermi LAT, and up to $\approx$ MJD 56800 it remained constant with no significant changes. Starting from $\approx$ MJD $56800$, the flux substantially increased up to a few times $10^{-7}{\rm photon\:cm^{2}\:s^{-1}}$ and remained so till $\approx$ MJD $57350$. The standard $\chi^2$ analysis revealed a highly variable \gray flux, where the probability of the flux to be constant is $p(\chi^2)<<5\%$. No strong variation of the \gray photon index is found during the time under consideration.\\
The active state identified above can be further investigated using denser time sampling. However, considering the relatively weak flux, the light curve will contain many upper limits preventing to make any conclusion. Therefore, a light curve generated with an adaptive binning method is used. In this method, the time bin widths are flexible and chosen to produce bins with constant flux uncertainty \citep{lott}. This method allows detailed investigation of flux changes in time, since at times of high fluxes, the time bins are narrower than during lower ones, therefore the rapid changes of the fluxes can be found. In order to reach the necessary relative flux uncertainty, the integral fluxes are computed above the optimal energies \citep{lott} ($E_0$ = 183.2 MeV in this case). Also, in order to improve the accuracy of the method, the flux of bright sources which lie close to 3C 120 have been taken into account. This is done by providing the parameters of confusing sources during the adaptive binning light curve calculations.\\
For 20\% adaptively binned intervals, light curve is generated for the energy range 100 MeV-300 GeV (lower panel in Fig. \ref{fg2}). As it is expected, initially it took a long time to reach the necessary 20 \% uncertainty. Indeed, the first bin contains the data from the start of the mission to MJD 56919.31(19 September 2014), amounting to more than 6 years. Afterwards, it took shorter time to reach the required uncertainty. The most dramatic increase in the \gray flux was observed on April 24, 2015. First, within 19.0 min the flux reached $(7.46\pm1.56)\times10^{-6}\;{\rm photon\:cm^{-2}\:s^{-1}}$ with $\Gamma=2.29\pm0.21$ and $11.2\sigma$ detection significance. Then for another 3.15 hours it was as high as $(4.71\pm0.92)\times10^{-6}\;{\rm photon\:cm^{-2}\:s^{-1}}$ with $\Gamma=1.97\pm0.14$ and $12.7\sigma$. Then the flux slowly decreased down to a few times $10^{-7}\;{\rm photon\:cm^{-2}\:s^{-1}}$ with the bin size varying within 10 to 35 days. The source was in an active state up to $\approx$ MJD 57300 and then turned again into its quiescent state, in which case the data should be accumulated for $\simeq$254 days.
\subsection{Spectral analysis}
In order to investigate the emission from 3C 120 in its quiet and active states, the \gray spectra were extracted from the following periods:
\begin{itemize}
\item[{\it 1)}] long quiescent states, namely, between MJD 54682.65 and MJD 56919.31.
\item[{\it 2)}] the active state after MJD 56919.31. The period overlaps with the Swift observations on MJD 56934.19, 56937.70 and 5638.50. Although the Swift observation lasted several thousands of seconds, in order to increase the \gray photon statistics, the Fermi LAT spectrum was extracted from 15 days (MJD 56919.31-56934.76), when the source showed a comparable flux level as inferred from the light curve obtained by an adaptive binning method.
\end{itemize}
The spectrum of 3C 120 was modeled as a power-law function ($dN/dE\sim N_{0}\:E^{-\Gamma}$) with the normalization and index considered as free parameters. In order to find the best matches between the spectral models and events, a binned likelihood analysis is performed with {\it gtlike} for the first period, while an unbinned analysis was applied for the second one. The spectral fitting results are summarized in Table \ref{tab:chandra:res} and the plot of the SEDs is shown in Fig. \ref{SEDnuclear}. During the flaring periods, the \gray flux increased nearly by an order of magnitude and the photon index hardened.
{\def\arraystretch{2}\tabcolsep=9pt
\begin{table}
\caption{Parameters of spectral analysis}
\label{tab:chandra:res}
\begin{center}
\renewcommand{\arraystretch}{1.5}
\begin{adjustbox}{width=0.48\textwidth}
\begin{tabular}{cccc}
\clineB{1-4}{3.5}
& Swift XRT & & \\
\clineB{1-4}{3.5}
Obsid & $\Gamma_{\rm X} $& \multicolumn{1}{c}{$\nu F_{\nu}$} & reduced $\chi^2$\\
& MJD & $\times10^{-11}{\rm erg\:cm^{-2}\:s^{-1}}$& \\
\hline
37594002 & $1.42\pm0.07$ & $4.27\pm0.18$ & 0.43 \\
37594004 & $1.53\pm0.08$ & $2.41\pm0.11$ & 0.75  \\
37594042 & $1.76\pm0.04$ & $6.73\pm0.15$ & 1.21  \\
37594048 & $1.72\pm0.04$ & $5.37\pm0.12$ & 1.05  \\
37594049 & $1.80\pm0.06$ & $3.73\pm0.13$ & 0.86 \\
\clineB{1-4}{3.5}
& Fermi LAT  & &\\
\clineB{1-4}{3.5}
date & Flux  & $\Gamma$  & TS \\
 & $10^{-8}{\rm photon\:cm^{-2}\:s^{-1}}$ & & \\
\hline
2008/08/04-2014/09/19 &$2.87\pm0.49$ & $2.79\pm0.08$ & 179.43 \\
2014/09/19-2014/10/04& $24.9\pm4.21$ & $2.57\pm0.16$ & 90.5 \\
\clineB{1-4}{3.5}
 & Chandra & &  \\
\clineB{1-4}{3.5}
Region&$\Gamma_{\rm X}$& $\nu F_{\nu}$ & reduced $\chi^2$\\
& &$\times10^{-14}{\rm erg\:cm^{-2}\:s^{-1}}$& \\
\clineB{1-4}{3.5}
k4 &$1.82 \pm 0.10$ & $15.97\pm2.3$ & 1.03 \\
k7 &$2.72\pm0.66 $& $1.85\pm0.82$ & 1.03 \\
s2 &$2.64 \pm1.26 $&$0.78\pm0.51$ & 0.81 \\
s3 &$2.14 \pm 0.28 $&$0.45\pm0.37$ & 0.89 \\
k25 inner &$1.63 \pm 0.22$&$3.89\pm1.42$ & 0.98\\
k25 outer &$1.62 \pm 0.11$& $12.16\pm1.9$ & 0.79 \\
k25 new &$1.80 \pm 0.19$&$6.28\pm1.67$ & 0.91 \\
\clineB{1-4}{3.5}
\end{tabular}
\end{adjustbox}
\end{center}
\end{table}
}
\subsection{Swift observations}
Swift satellite \citep{Gehrels} observed 3C 120 in its \gray quiescent and active states. As the X-ray flux varies as well (see $www.bu.edu/blazars/VLBA\_ GLAST/3c120.html$), we have selected the observations made on MJD 55252.70 and MJD 55800.25 when the X-ray flux also was low. During the \gray active state 3C 120 was observed only three times, on MJD 56934.19, MJD 56937.70 and MJD 56938.50. The data from two of the instruments on board Swift, the UltraViolet and Optical Telescope (UVOT) and the X-Ray Telescope (XRT), have been used in the analysis.
\subsection{Swift XRT}
The Swift-XRT observations were made in the photon counting (PC) (Obsid 37594002, 37594004) and windowed timing (WT) (Obsid 37594042, 37594048, 37594049) modes. The data were analyzed using the {\it XRTDAS} software package (v.3.3.0) distributed by HEASARC as part of the HEASoft package (v.6.21). The source spectrum region was defined as a circle with a radius of 30 pixels ($\sim 71''$) at the center of the source, while the background region - as an annulus centered at the source with its inner and outer radii being 80 ($\sim 190''$) and 120 pixels ($\sim 280''$), respectively. For the PC-mode observation 37594004, the count rate was above 0.5 count/s, being affected by the piling up in the inner part of the PSF. This effect was removed by excluding the events within a 4-pixel radius circle centered on the source position. Then, using {\it xrtmkarf} task, ancillary response files were generated by applying corrections for the PSF losses and CCD defects.\\
The spectrum was rebinned to have at least 20 counts per bin, ignoring the channels with energy below 0.5 keV, and fitted using {\it XSPEC} v12.9.1a. The results of the fit are given in Table \ref{tab:chandra:res} and the corresponding spectra are shown in Fig. \ref{SEDnuclear}. The 0.5-6.0 keV spectrum is well fitted by an absorbed power-law model with column density $N_{H}=1.06\times 10^{21}\:{\rm cm^{-2}}$. Although the X-ray flux did not increase significantly (the highest flux of $F_{0.5-6\:{\rm keV}}\simeq(6.73\pm 0.15)\times10^{-11}\:{\rm erg\:cm^{-2}\:s^{-1}}$ observed on October 4, 2014, exceeds the lowest one $\sim2.8$ times), the X-ray photon index softened, changing in the range of $\Gamma_{\rm x}=(1.76-1.80)$ during the bright \gray periods.
\subsection{Swift UVOT}
\begin{figure*}
   \centering
  \includegraphics[width=0.48 \textwidth]{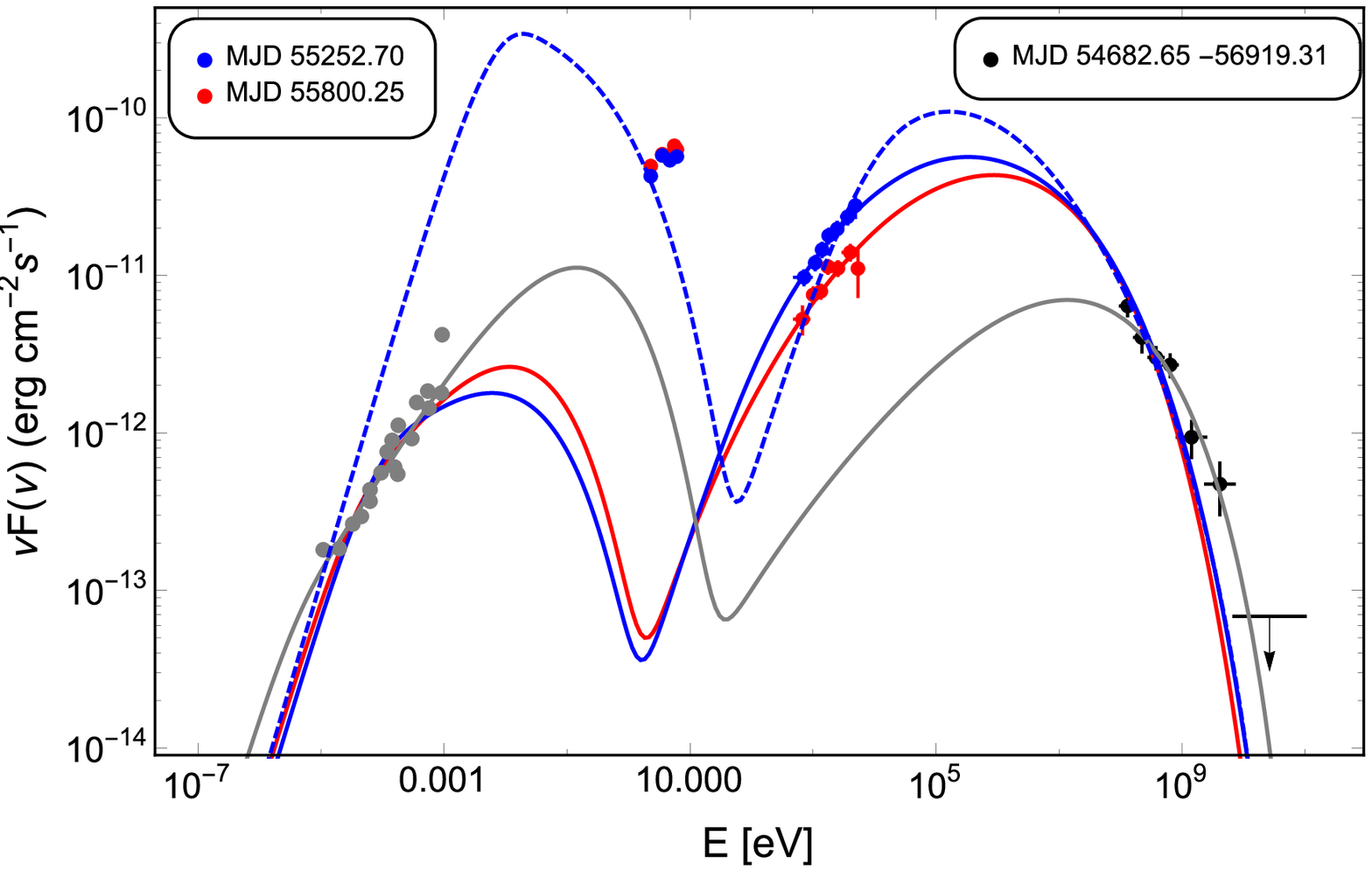}
  \includegraphics[width=0.48 \textwidth]{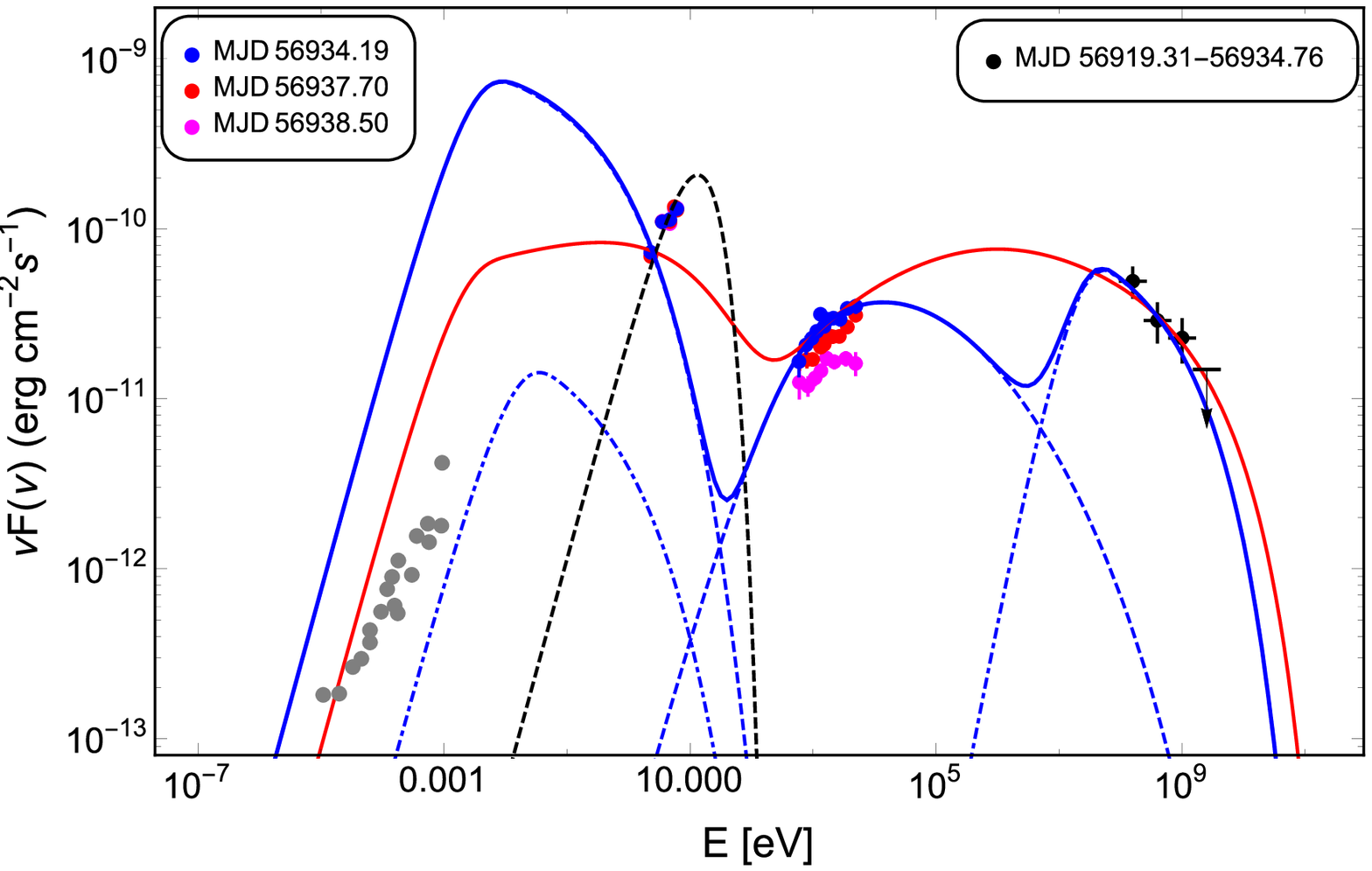}
   \caption{The broadband SED of 3C 120 core emission for quiescent (left) and flaring (right) states. {\it Left panel:} The blue and red solid lines are the synchrotron/SSC model fitting for two different X-ray fluxes, taking into account the radio data and assuming Swift UVOT data are upper limits. Instead the dashed blue line is calculated assuming optical/UV emission is also produced by the jet. {\it Right panel:} The SED in flaring state fitting with one-zone synchrotron/SSC (red solid line) and two zone SSC+ EIC (blue solid line) models. The model parameters are presented in Table 2.}
    \label{SEDnuclear}
\end{figure*}

In the analysis of the Swift UVOT data, the source counts were extracted from an aperture of $5.0''$ radius around the source. The background counts were taken from the neighboring circular region having a radius of $20''$. The magnitudes were computed using the {\it uvotsource} tool (HEASOFT v6.21), corrected for extinction according to \citet{roming} using $E(B-V)=0.256$ from \citet{schlafly} and zero points from \citet{breeveld}, converted to fluxes following \citet{poole}. The corresponding spectra are shown in Fig. \ref{SEDnuclear}. The optical-UV data points harden during the flaring periods.
\section{Chandra observations of the knots}\label{sec3}
In the public archive we found 5 observations (ObsId 3015, 16221, 17564, 17565, 17576) of 3C 120 with Chandra telescope, the overall observation time being 251.86 ksec. We applied the standard data reduction procedure, using the Chandra Interactive Analysis of Observations (CIAO) 4.8 with Chandra Calibration Database (CALDB) version 4.7.2. We checked for flaring background events and did not find any significant flares. The readout streaks were removed for each observation and the events files were then re-projected to a single physical coordinate system (using observation 16221 as a reference). Also, to reduce the uncertainties caused by the position offsets of different observations, we made astrometric corrections, following the thread from $www.cxc.harvard.edu/ciao/threads/reproject\_ aspect/$. For each knot, the source region and the background are selected based on the position and shape given in Table 3 and Fig. 3 of \citet{hariss04}.\\
We extracted the spectra and created weighted response files for each observation, using the {\it specextract} script. The knots spectra were rebinned using a different count threshold depending on the total number of counts, and fitted in the 0.5-10 keV energy range using a power-law with the Galactic absorption model (column density fixed at $N_{H}=1.06\times10^{21}\:{\rm cm^{-2}}$), where the index and the normalization are allowed to vary freely. The spectral fit was done with {\it Sherpa} using {\it levmar} optimization method and {\it chi2datavar} statistics. \\
The fitted spectral parameters are summarized in Table \ref{tab:chandra:res}. The X-ray emission from the core is heavily saturated making it impossible to study the innermost parts of the jet. However, the nearby k4 knot’s emission is well resolved from the core, the X-ray spectral index being $1.82\pm0.1$ and the flux being $(1.60\pm0.23)\times10^{-13}\:{\rm erg\:cm^{-2}\:s^{-1}}$. The net counts from k7, s2 and s3 knots are relatively low $<50$ (as compared with $>100$ from other knots), not enough for a detailed spectral fitting. The fitting resulted in a steep X-ray slope ($>2.0$) and a relatively faint X-ray flux ($\leq10^{-14}\:{\rm erg\:cm^{-2}\:s^{-1}}$). Because of low statistics and estimated large uncertainties, we did not consider them further. Following \citet{hariss04} we also sub-divide k25 into inner, outer and new regions which have different properties in the radio band (the inner refers to the upstream edge and the outer refers to the western edge, see Fig. 3 of \citet{hariss04}). The X-ray emission from all 3 regions has harder X-ray emission spectra ($\leq 1.8$) with the X-ray flux varying within $(3.89-12.16)\times10^{-14}\:{\rm erg\:cm^{-2}\:s^{-1}}$. It is interesting to note that the flux of k25 outer is at the same level as that of the bright k4 but with a significantly harder X-ray spectral index of $\Gamma_{\rm X}=1.62\pm0.11$. The knot SEDs shown in Fig. \ref{knot_spectra} have been calculated by {\it sample\_energy\_flux} in Sherpa.
\section{Modeling the spectral energy distributions}\label{sec4}
\begin{table*}
\footnotesize
\renewcommand{\arraystretch}{1.7}
\begin{center}
\caption{Parameters obtained from the fit of the emission from the inner jet of 3C 120 during quiescent and flaring states.}
\label{table_fit}
\begin{tabular}{lccccc}
\hline
& Parameter & SSC(blue-dashed)  & SSC (red) & SSC & SSC+EIC \\
\hline
Doppler factor & $\delta$ & 4 & 4 & 6 & 4(6)\\
Normalization of electron distribution & $N_0^\prime\times10^{50}\:{\rm eV^{-1}}$ & $15292.18_{-10357.06}^{+34383.07}$ & $16.77_{-14.02}^{+176.69}$ & $1.96_{-1.13}^{+2.75}$    & $1713.92^{+19.88}_{-18.69}$($817.55_{-117.96}^{+118.64}$) \\
Electron spectral index & $\alpha$ & $3.12_{-0.16}^{+0.15}$ & $1.85_{-0.22}^{+0.31}$ & $2.79_{-0.13}^{+0.16}$ & $3.12\pm0.22$($3.24^{+0.89}_{-0.80}$) \\
Minimum electron energy & $E^\prime_{\rm min}$ (MeV) & $354.51_{-24.55}^{+27.20}$ & $228.18_{-120.10}^{+92.19}$ & $67.57_{-20.65}^{+18.03}$ & $117.23^{+11.80}_{-12.64}$($514.51^{+569.19}_{-352.58}$) \\
Cut off electron energy &$E^\prime_{\rm cut}$ (GeV) & $3.21_{-0.46}^{+0.60}$ & $4.61_{-0.77}^{+1.63}$ & $6.32_{-1.48}^{+2.93}$ & $1.68^{+0.42}_{-0.41}$($4.05^{+5.37}_{-2.10}$) \\
Maximum electron energy & $E^\prime_{\rm max}$ (TeV) & $2.30_{-1.54}^{+3.84}$ & $1.90_{-1.26}^{+3.02}$ & $1.83_{-1.27}^{+2.19}$  & $10.71^{+4.64}_{-6.56}$($54.56^{+75.36}_{-40.27}$) \\
Magnetic field & B [G] & $0.16_{-0.007}^{+0.006}$ & $0.0023_{-0.00018}^{+0.00025}$ &  $0.86^{+0.11}_{-0.09}$   & $0.63\pm0.12$($0.11^{+0.11}_{-0.08}$)\\
Electron energy density & $U_{e}({\rm erg\:cm^{-3}})$& $1.02\times10^{-3}$ & $1.99\times10^{-2}$ &  $0.14$   & $4.39\times10^{-4}$($0.25$)\\
Jet power in magnetic field & $L_{B}\times10^{44}$ erg s$^{-1}$ & $2.58$ & $4.98\times10^{-4}$ & 0.22  &  $38.34(0.0034)$ \\
Jet power in electrons & $L_{e}\times10^{44}$ erg s$^{-1}$ & $2.46$ & $48.00$ & 1.09 & $1.06(2.02)$\\
\end{tabular}
\end{center}
\end{table*}
\subsection{The core region}
The broadband SEDs of 3C 120 core’s emission in its quiescent and flaring states are shown in Fig. \ref{SEDnuclear} with the radio data (gray) from \citep{giommi,tanaka3c120} where the data from the period corresponding to the \gray quiescent state of 3C 120 is analyzed. As in the previous studies \citep{giommi,sahak3c120,tanaka3c120,sikora16} the SEDs hint at the existence of two nonthermal emission peaks in the IR/optical/UV and HE \gray bands. The UVOT data revealed a rather hard optical-UV spectrum which indicates that perhaps direct thermal emission from the accretion disk was being observed \citep{ghisellini09,ghisellini009,ghisellini11}.\\
Taking into account the results of the previous studies of other Fermi LAT-observed radio galaxies \citep{abdom87,abdo2010a,sahak3c120,tanaka3c120}, the multiwavelength emission of 3C 120 is modeled using the synchrotron/Synchrotron Self-Compton (SSC) \citep{maraschi,bloom,ghisellini} and/or External Inverse-Compton (EIC) \citep{sikora} models. The radio through optical emission is due to the synchrotron emission of energetic electrons in the homogeneous, randomly oriented magnetic field, while the X-ray to HE \gray emission is due to the inverse Compton scattering of the same synchrotron photons or photons of external origin.\\
In the flaring state (right panel of Fig. \ref{SEDnuclear}) the X-ray flux moderately increased and the spectrum softened, while the HE \gray flux increased and its spectrum shifted to higher energies. Within the synchrotron/SSC or EIC scenarios such modifications can be explained by means of introducing changes in the electron acceleration, increase of the comoving radius and bulk Lorentz factor \citep{paggi} or due to the contribution from external photons. Here we discuss the following possibilities: {\it (i)} in the quiescent state the jet energy dissipation occurs close to the central black hole and the dominant mechanism is the synchrotron/SSC emission, {\it (ii)} in the flaring period again the dominant processes is SSC, although the emission region has a higher bulk Lorentz factor, and {\it (iii)} the optical/UV/X-ray emission is due to the synchrotron/SSC emission from an extended and slow-moving region, while the HE \grays come from a compact and fast-moving region, where EIC is dominating. This is similar to the scenario adopted by \citet{tavecchio11} to explain the very fast VHE \gray variations and the hard GeV spectrum of PKS 1222+216. The choice of this scenario is justified since strong changes are observed only in the \gray band.\\
The emission region (the "blob") is assumed to be a sphere with a radius of $R$ which carries a magnetic field with an intensity of $B$ and a population of relativistic electrons which have a power-law with an exponential cut-off energy distribution expected from shock acceleration theories \citep{inou}:
\begin{equation}
N^{\prime}_{\rm e}(E^{\prime}_{\rm e})= N^{\prime}_{0}\:\left( E^{\prime}_{e}/m_{e}\:c^2\right)^{-\alpha}\:Exp[-E^{\prime}_{\rm e}/E^{\prime}_{\rm cut}]
\label{BPL}
\end{equation}
for $E^{\prime}_{\rm min}\leq ^{\prime}E_{\rm e}\leq E^{\prime}_{\rm max}$ where $E^{\prime}_{\rm min}$ and $E^{\prime}_{\rm max}$ are the minimum and maximum electron energies, respectively. The total electron energy $U_{\rm e}=\int_{\rm E^{\prime}_{\rm min}}^{\rm E^{\prime}_{\rm max}}E^{\prime}_{e} N^{\prime}_e(E^{\prime}_{e})dE^{\prime}_e$ is defined by $N^{\prime}_{0}$, $\alpha$ is the electron spectral index and $E^{\prime}_{\rm cut}$ is cut-off energy.\\
Since the blob moves along the jet with a bulk Lorentz factor of $\Gamma_{\rm bulk}$, the radiation will be amplified by a relativistic Doppler factor of $\delta=1/\Gamma_{\rm bulk}(1-\beta cos(\theta_{\rm obs}))$, where $\theta_{\rm obs}$ is the angle between the bulk velocity and the line of sight. For 3C 120, the averaged bulk Lorentz factor has been estimated to be $\Gamma_{\rm bulk}=5.3\pm1.2$ \citep{jorsted,casadio}, while different mean values for $\theta_{\rm obs}$ were obtained in VLBI observations; $\theta_{\rm obs}$ varies from $9.7^\circ$ \citep{hovatta} to $20.5^\circ$ \citep{jorsted}. So, we assumed $\delta=4$ (e.g., $\theta_{\rm obs}\sim[15^\circ-20^\circ]$) and $\delta\approx6$ (e.g., $\theta_{\rm obs}\sim9.7^\circ$) for the quiescent and flaring states, respectively. In the quiescent state, since no significant \gray variability is observed (or it varies in a long period) most likely the emission is produced in a large region for which we assume $R\approx4\times10^{17}{\rm cm}\sim0.1\:{\rm pc}$. Instead, in the active state the fast \gray flares in day/sub-day scales indicate that the emitting region size should be $R/\delta\leq c\times t\times \delta=1.56\times10^{16}\;(t/{\rm 1\:day})\;(\delta/6)\:{\rm cm}$.\\
In the flaring state, the inverse Compton scattering of external photons either reflected from the BLR \citep{sikora} or from the hot dusty torus \citep{blazejowski,ghismarasch} can contribute to the emission in the \gray band. For any reasonable assumption about the jet opening angle ($\theta=0.1^{\circ}$) and Doppler boosting factor ($\delta=6$) the \gray emission region is at the distance $\sim R/\theta\approx 2.3\:{\rm pc}$ well beyond the radius of BLR $(5.9-7.4)\times10^{16}\:{\rm cm}$ determined from reverberation mapping \citep{pozo}. In this case the dominant external photon field is the IR radiation from the hot dusty torus which, as we assume, has a blackbody spectrum with a luminosity of $L_{\rm IR} = \eta L_{\rm disk}$ ($\eta = 0.6$) \citep{ghisellini009} and these photons are filling a volume that for simplicity is approximated as a spherical shell with a radius of $R_{\rm IR} = 3.54\times10^{18} (L_{\rm disk}/10^{45})^{0.5} {\rm cm}$ \citep{nenkova}. The accretion disk luminosity was estimated using the Swift UVOT data points observed during the flaring periods. Reproducing the UVOT fluxes with the Shakura-Sunyaev disk spectrum \citep{shakura} fixing the peak energy at $\sim10\:{\rm eV}$, we obtained that the disk luminosity $L_{disk}=1.2\times10^{45}\:{\rm erg\:s^{-1}}$ (see Fig.\ref{SEDnuclear} black dashed line), which is close to the value obtained in \citet{sikora16}.
\begin{figure*}
   \centering
  \includegraphics[width=0.48 \textwidth]{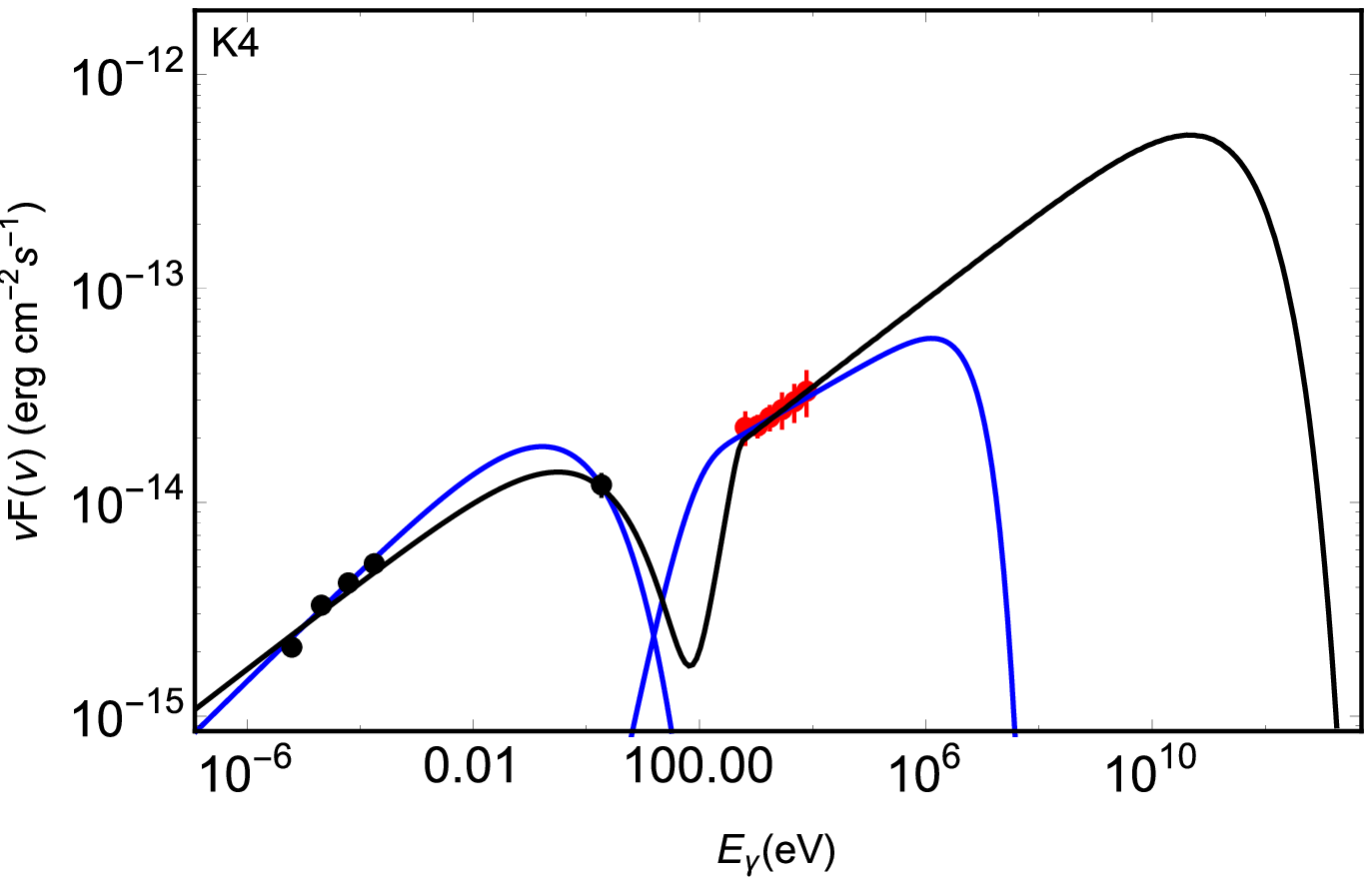}
   \includegraphics[width=0.48 \textwidth]{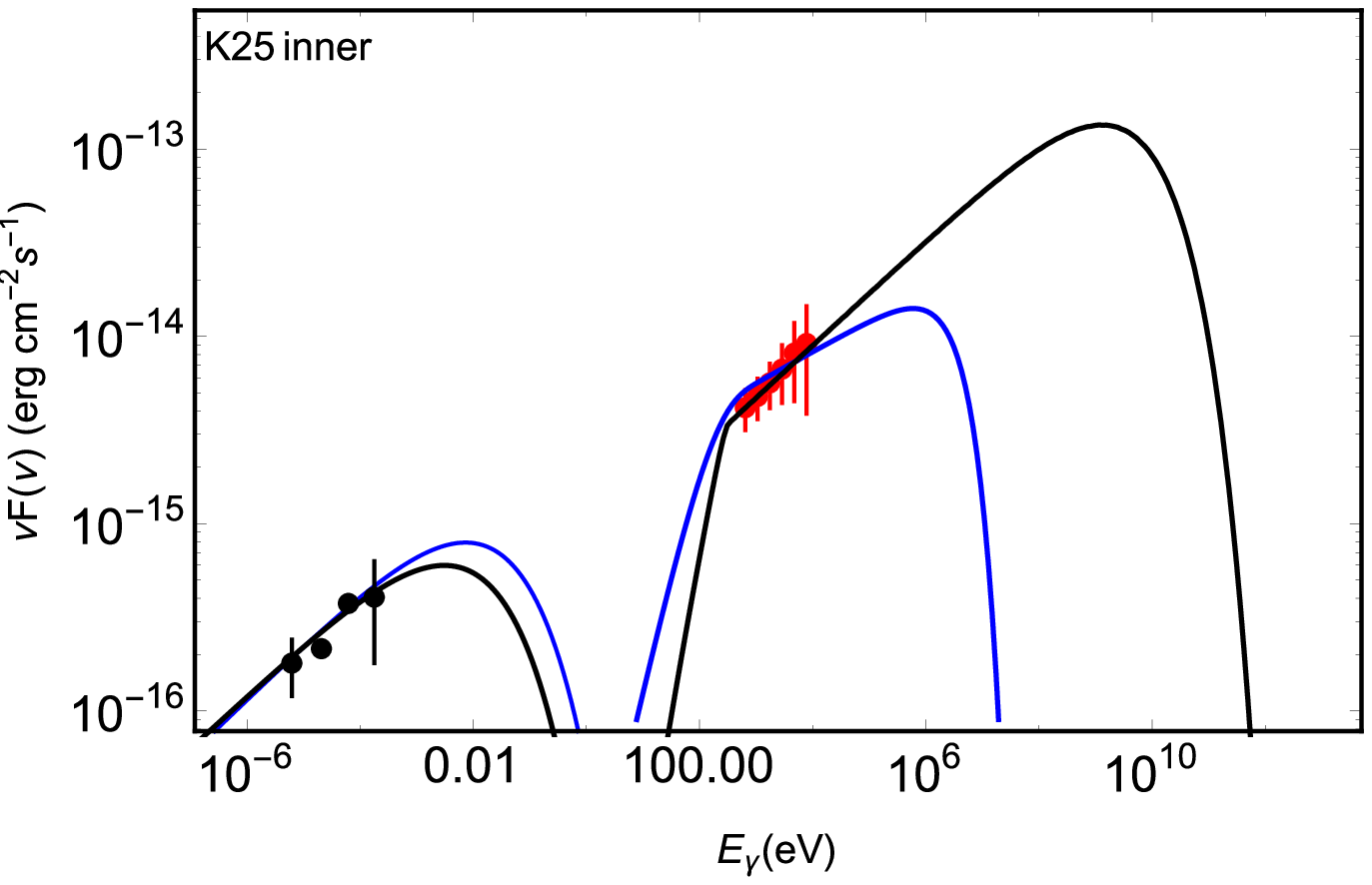}\\
   \includegraphics[width=0.48 \textwidth]{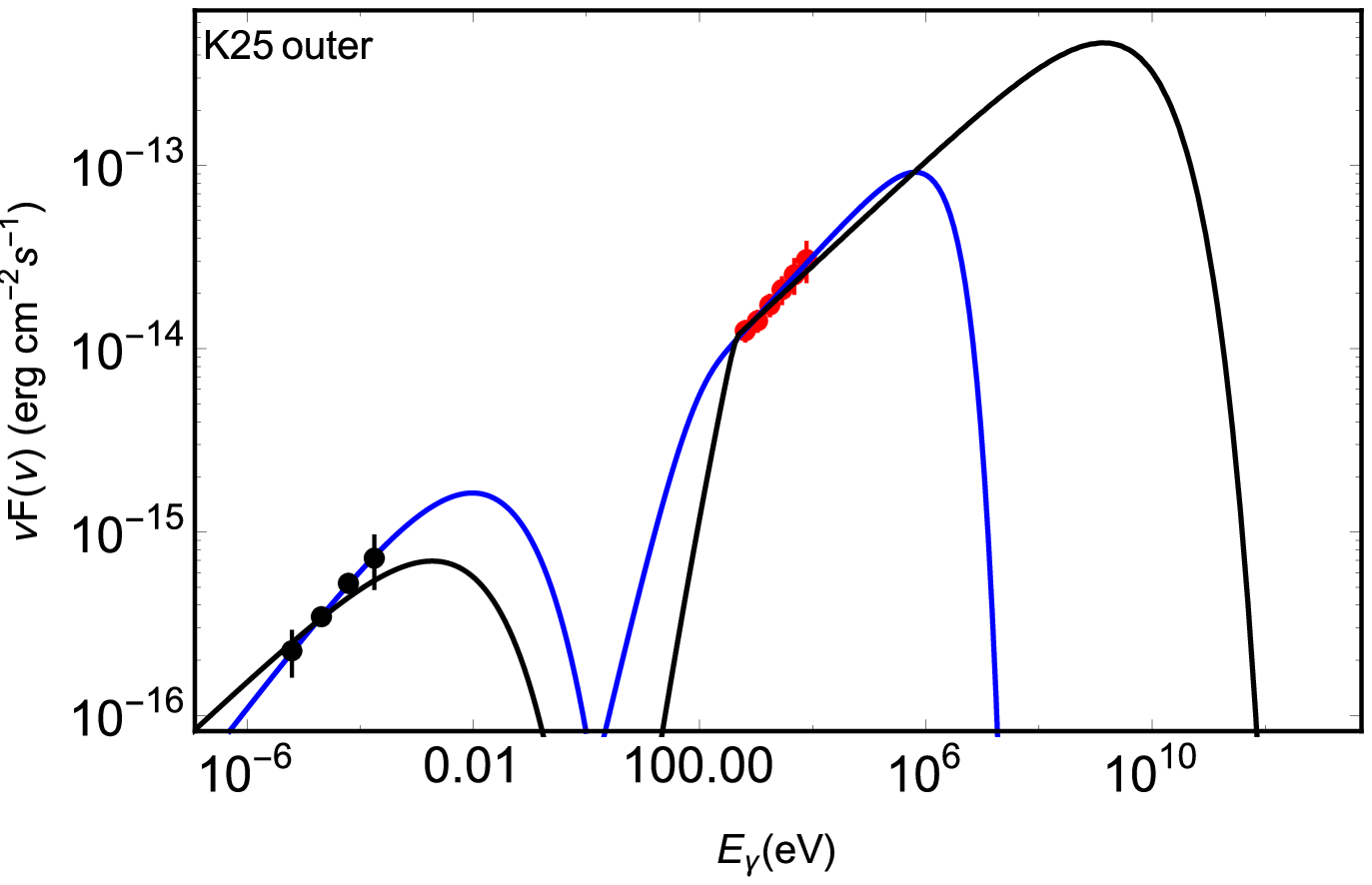}
   \includegraphics[width=0.48 \textwidth]{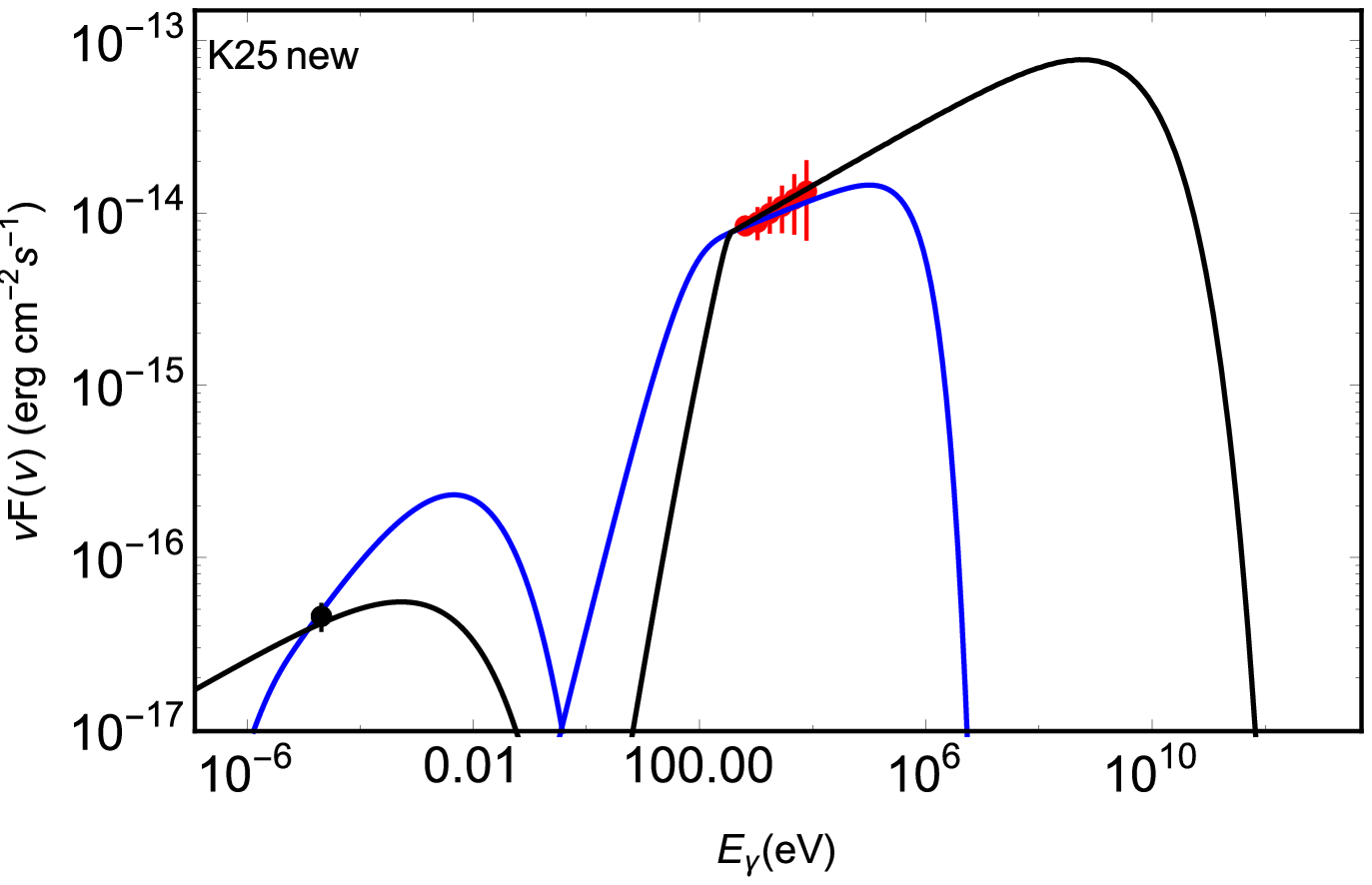}
   \caption{The SED of knots. The radio-to-optical data (black points) are from previous studies and the X-ray data (red points) are derived in this paper. Black lines are the IC/CMB model calculated for $\delta=10$, and blue lines are the fit by the two-component synchrotron model.}
    \label{knot_spectra}
\end{figure*}
\subsection{The large-scale jet emission}
We assembled the 3C 120 knots’ SEDs from the radio to X-ray bands, using the radio-to-optical data from \citet{hariss04} (black points in Fig. \ref{knot_spectra}) and the X-ray data obtained in Section \ref{sec3} (red points in Fig. \ref{knot_spectra}). The X-ray flux is well above the extrapolation from the radio-to-optical spectra and it hardens; this is more evident for the k25\_new where $L_{\rm X}/L_{\rm rad}\simeq 250$. This confirms the previous findings that two different components are necessary to explain the radio-to-optical and X-ray emission from the knots of 3C 120.\\
The detected highly polarized emission led to a conclusion that the radio-to-optical emission from the knots is of a synchrotron origin. The radiative mechanism usually considered to explain the origin of the X-ray emission is either the synchrotron emission from a second, much more energetic population of electrons (e.g., \citep{jester,hariss04,marshal,roser}) or the inverse Compton scattering on various possible sources of soft photons, including the synchrotron photons (SSC; e.g., \citep{hardcastle2002}) and the cosmic microwave background photons (IC/CMB: e.g., \citep{tavecchiocmb}). Most naturally, the X-ray emission could originate from IC scattering of synchrotron photons with a density of $U_{\rm syn}=L_{\rm radio}/4 \pi R_{\rm k4}^2\:c\approx3.2\times10^{-14}\;{\rm erg\:cm^{-3}}$ where $L_{\rm radio}\simeq2.0\times10^{40}\;{\rm erg\:s^{-1}}$ is the radio luminosity of k4 and $R_{\rm k4}\approx0.42\;{\rm kpc}$ is the knot size. When comparing the radio and X-ray data it becomes clear that $L_{\rm SSC}\geq L_{\rm syn}$ ($U_{\rm B}\leq U_{\rm syn})$ which is satisfied only if $B\leq0.8\:{\rm \mu G}$, which is in contradiction with the averaged value of $\geq10\:{\rm \mu G}$ usually estimated for the knots. In such a weak magnetic field, the observed radio luminosity can be accounted for only if the particle energy ($U_{\rm e}$) strongly dominates over the magnetic field thus contravening the equipartition condition. Because of this we only consider (1) the inverse Compton scattering of CMB photons and (2) synchrotron emission from a second, much more energetic population of electrons. In the IC/CMB scenario, it is assumed that the emitting region moves with a relativistic bulk Lorentz factor of $\Gamma_{\rm bulk}$ in order to predict a larger X-ray luminosity, since in the jet frame the energy density of CMB photons will be enhanced by a factor of $\Gamma_{\rm bulk}^{2}$. The condition of $L_{\rm X-ray}>L_{\rm radio}$ is satisfied only if $\delta$ is at least $10$. In contrast, if the X-ray emission is produced by synchrotron radiation from a second population of relativistic electrons with very high maximum energy, it is not required to have a highly relativistic jet, and we assume $\delta=1$. Since the electrons with high maximum energy would quickly cool down, most likely they had been produced in a separate episode of acceleration, which occurred more recently. Moreover, the second population of electrons can be produced in different parts of the knot [e.g., \citep{ostrowski,stawarz}]. We do not go much into details, but assume that there are two different electron populations responsible for the emission.\\
In the IC/CMB scenario, the underlying electron distribution is given by Eq. \ref{BPL} with the cutoff energy allowed to vary only for k4, where the optical flux at $\sim10^{14}$ Hz allows to constrain the HE tail of synchrotron emission, while for the other knots it is fixed at $E_{\rm cut}=100$ GeV due to the lack of data. In addition, a power-law distribution in the form of $N_{\rm PL}\sim E_{\rm e}^{-\alpha_{\rm PL}}$ is added to model the X-ray spectrum in the two-component synchrotron scenario. Since the data are not enough to constrain $E_{max,PL}$, an artificial HE limit of $E_{\rm PL, max}=1$ PeV has been introduced, whereas $E_{\rm PL, min}$ is left as a free parameter only ensuring that the flux from the second component does not exceed the first one. In our calculations we used the shapes and sizes of the knots as provided in \citet{hariss04}. To make the deviation from the equipartition condition as less as possible, we define $\eta=U_{\rm e}/U_{\rm B}$, which is used as a free parameter with $U_{\rm e}$ during the fit. This will allow to find the optimum value for $\eta$ when $\eta=1$ does not give satisfactory representation of the data. In the two-component synchrotron model, we fix $\eta=1$ and perform fitting of the radio-to-optical data. Then fixing this magnetic field, the X-ray data are fitted with the second component.
\subsection{Fitting technique}
In order to constrain the model free parameters we have modified the {\it naima} package \citep{zabalza} which derives the best-fit and uncertainty distributions of spectral model parameters through MCMC sampling of their likelihood distributions. The prior likelihood, our prior knowledge of the probability distribution of a given model parameter and the data likelihood functions are passed onto the emcee sampler function for an affine-invariant Markov Chain Monte Carlo run. In the parameter sampling, the following expected ranges are considered: $1.5\leq(\alpha, \alpha_{\rm PL})\leq10$, $0.511\:{\rm MeV}\leq E^\prime_{cut}\leq1\:{\rm TeV}$ and $N_0$ and $B$ are defined as positive parameters. The synchrotron emission is calculated using the parameterization of the emissivity function of synchrotron radiation in random magnetic fields presented in \citet{aharsyn} while the IC emission is computed based on the monochromatic differential cross section of \citet{aharatoy}.
\begin{table*}
\footnotesize
\renewcommand{\arraystretch}{1.7}
\begin{center}
\caption{The parameters derived from the modeling of the knots’ SEDs.}
\label{table_fit_knot}
\begin{tabular}{lcccc}
\hline
Parameter & k4  & k25 inner & k25 outer & k25 new \\
\hline
$U_{e}^{\prime}(U_{e})[{\rm erg\:cm^{-3}}]\times10^{-12}$ & $111.8_{-9.91}^{+16.49}(531.52_{-84.76}^{+110.7})$& $20.65_{-3.47}^{+9.95}(102.91_{-45.22}^{+82.45})$& $30.25_{-2.73}^{+6.21}(48.56_{-13.47}^{+26.11})$& $7.43_{-1.43}^{+2.89}(4.09_{-0.63}^{+1.27})$\\
$\alpha$ & $2.59\pm0.03(2.48\pm0.06)$ & $2.44\pm0.1(2.41^{+0.22}_{-0.26})$ & $2.42\pm0.04(2.20_{-0.15}^{+0.17})$ & $2.62_{-0.10}^{+0.14}(2)$\\
$E_{\rm min}^{\prime}(E_{\rm min})[MeV]$ & $24.01_{-5.44}^{+3.62}(4.40_{-2.31}^{+2.92})$& $18.2\pm10.6(3.39_{-1.90}^{+3.57})$& $22.23_{-8.43}^{+5.53}(4.04_{-2.92}^{+8.48})$& $19.1_{-8.57}^{+7.11}(1445.67_{-1198.91}^{+3852.81})$\\
$B^{\prime}(B)$ [$\mu$G] & $5.61(115.6)$ & $2.35(50.85)$ &  $1.36(34.94)$   & $0.93(10.14)$ \\
$\eta $ & $89.39_{-12.75}^{+7.48}$ & $94.04_{-43.7}^{+49.8}$ &  $410.86_{-104.94}^{+64.28}$   & $214.3_{-124.63}^{+162.67}$ \\
$U_{e,p}({\rm erg\:cm^{-3}})\times10^{-14}$ & $6.01_{-2.38}^{+5.85}$& $2.36_{-1.19}^{+2.59}$& $7.22_{-1.93}^{+3.01}$& $13.22_{-5.62}^{+15.7}$\\
$\alpha_{\rm p}$ & $2.69_{-0.18}^{+0.22}$ & $2.66_{-0.54}^{+1.15}$ & $2.32^{+0.17}_{-0.18}$ & $2.73_{-0.24}^{+0.29}$\\
$E_{\rm min, p}[TeV]$ & $2.83_{-1.73}^{+3.22}$& $6.89_{-5.17}^{+9.83}$& $5.31_{-3.16}^{+5.65}$& $8.88_{-6.78}^{+10.24}$\\
$L_{B}\times10^{42}$ & $6.48(128)$& $71.26(333.7)$& $31.98(211)$& $0.67(3.69)$\\
$L_{e}\times10^{44}$ & $5.81(1.28)$& $66.98(3.34)$& $131.45(2.11)$& $1.44(0.04)$\\
\end{tabular}
\end{center}
\end{table*}
\section{Results and discussion}\label{sec5}
In this paper, the multiwavelength emission from the 3C 120 core is investigated using the Swift XRT/UVOT and Fermi LAT data. Quiescent and flaring states are identified and their modeling allowed to investigate the jet properties and physical processes that take place in the core where, most likely, the jet is formed. On the other hand, the jet properties are also estimated at large distances from the core using the Chandra X-ray data.\\
The adaptively binned \gray light curve showed that before MJD 56900 and after MJD 57300, the source was in a quiescent state characterized by a relatively faint \gray emission with the flux and the photon index consistent with the previously reported values. Then, from MJD 56900 to MJD 57300, most of the time, the source was in an effective \gray emitting state with rapid \gray flares. During the bright periods, the \gray photon index hardened and corresponds to $\approx2.3$ and $\approx2.0$. Two strong events with $(7.46\pm1.56)\times10^{-6}\;{\rm photon\:cm^{-2}\:s^{-1}}$ and $(4.71\pm0.92)\times10^{-6}\;{\rm photon\:cm^{-2}\:s^{-1}}$ within accordingly 19.0 min and 3.15 hours were detected on April 24, 2015, which are the highest fluxes detected so far from 3C 120. 
At a distance of 144.9 Mpc, these correspond to an isotropic \gray luminosity of $(1.20-1.66)\times10^{46}\:{\rm erg\:s^{-1}}$. The same value estimated for the quiescent state (the first bin in Fig. \ref{fg2} red data) is $2.85\times10^{43}\:{\rm erg\:s^{-1}}$. Yet, assuming $\delta=6$, the total power emitted in the \gray band in the proper frame of the jet would be $L_{{\rm em,}\gamma}=L_{\gamma}/2\:\delta^2=(1.67-2.31)\times10^{44}\:{\rm erg\:s^{-1}}$ during the peak flux and $L_{{\rm em,}\gamma}=8.9\times10^{41}\:{\rm erg\:s^{-1}}$ in a quiescent state (assuming $\delta=4$). Thus, during the peak emission, the energy released in the \gray band corresponds to large fraction of Eddington luminosity ($L_{\rm Edd}=6.5\times10^{45}\:{\rm erg\:s^{-1}}$ for the black hole mass of $5.5\times10^{7}M_\odot$ \citep{peterson04}) while it is a small fraction ($\sim1.4\times10^{-4}$) in the quiescent state.\\
Usually the radio galaxies have a luminosity of $\leq10^{44}\:{\rm erg\:s^{-1}}$ \citep{abdomisaligned,ackerman2015}, and the peak \gray apparent luminosity of $(1.20-1.66)\times10^{46}\:{\rm erg\:s^{-1}}$ is unusual, more characteristic for BL Lac objects. Such a strong \gray output observed from 3C 120 is not surprising as the jet inclination angle is relatively small as compared with other radio galaxies.\\
In the X-ray band, the average flux in the 0.5-10.0 keV range is around $(2.4-4.3)\times10^{-11}\:{\rm erg\:cm^{-2}\:s^{-1}}$ in the quiet state and $(5.4-6.7)\times10^{-11}\:{\rm erg\:cm^{-2}\:s^{-1}}$ in October 4-7, 2014 (active state). When the lowest and highest fluxes from Table \ref{tab:chandra:res} are compared, a nearly 2.8 times increase of the X-ray flux is found, but its amplitude is lower as compared with the substantial increase in the \gray band. During the considered periods, the source spectra were always hard, $\Gamma_{\rm X}<2.0$. At bright \gray flares, the X-ray photon index softened ($1.72-1.80$) as compared with the hard photon index of $1.42-1.53$ in a quiescent state. The small change in the X-ray flux level and the photon index softening might indicate that different mechanisms are contributing to the acceleration and/or cooling of electrons which modifies the power-law index and the minimum energy of underlying electrons.\\
The data accumulated during several observations of 3C 120 allowed to resolve and study the X-ray emission from the large-scale jet of 3C 120. In particular, the counts from k4 and k25 are high enough for detailed spectral analysis, resulting in a flux of $\leq1.60\times10^{-13}\:{\rm erg\:cm^{-2}\:s^{-1}}$ with hard X-ray photon indices $\leq1.8$ which imply that most of the energy is released above 10 keV. It is interesting to note that the X-ray flux of $(1.22\pm0.19)\times10^{-13}\:{\rm erg\:cm^{-2}\:s^{-1}}$ from k25 outer located at $\sim16$ kpc from the core is of the order of the flux from the nearby k4 at $\sim2.5$ kpc however with a much harder, $\Gamma_{\rm X}=1.62\pm0.11$, X-ray photon index.
\subsection{The origin of emission from the inner Jet}
The broadband emission modeling results obtained in the quiescent and flaring states are shown in Fig. \ref{SEDnuclear} with the corresponding parameters in Table \ref{table_fit}.
In the quiescent state, the X-ray spectra have different photon indicies, $\Gamma_{\rm X}=1.42\pm0.07$ and $1.53\pm0.08$, this is why both spectra have been considered during the fit. In the fit we also included the archival radio data from the observations in the period when the source was in the quiescent state. The radio emission can be produced from the low-energy electrons which are accumulated for longer periods, so the the radio flux should not exceed the presented limit. When both radio and optical/UV data are considered, in order to have model which predicts emission below the radio flux, a larger value of $E^{\prime}_{\rm min}$ is required. However, the increase of $E^{\prime}_{\rm min}$ would also affect the flux predicted by SSC in the X-ray band; for example, the blue dashed line in Fig. \ref{SEDnuclear} illustrates the model for $E^{\prime}_{\rm min}\simeq354.51\pm25.91$ MeV (summing the errors in quadrature), beyond which the model predicts either a high radio flux or a low X-ray flux. The magnetic field is $B=0.16$ G with an energy density of $U_{\rm B}=1.07\times10^{-3}\;{\rm erg\:cm^{-3}}$, slightly higher than that of the electrons, $U_{\rm e}=1.02\times10^{-3}\;{\rm erg\:cm^{-3}}$. This magnetic field energy density should be considered as an upper limit, since the Swift UVOT data may represent the direct thermal disc emission, so, in principle, the synchrotron component can be much lower. Thus, in the second modeling, we assume that the low-energy component is only defined by the radio data (blue and red solid lines in Fig. \ref{SEDnuclear}). In this case the underlying electron distribution is characterized by a harder power-law index ($\alpha=2.22\pm0.19$ and $\alpha=1.85\pm0.27$ for blue and red solid lines respectively) and higher cut-off energy $E^{\prime}_{\rm cut}=4.61\pm1.27$. The magnetic field is significantly lower, $B=2.3\times10^{-3}$ G, and the jet should be strongly particle-dominated to have the peak flux of the HE component exceeding that of the low energy one. This particle dominance can be minimized assuming that the X-ray emission is of a different origin (e.g., from another blob or from thermal Comptonization near the disk). When the optical/UV and X-ray data are assumed as upper limits in the fit, a lower flux from SSC emission is expected (gray line in Fig. \ref{SEDnuclear} [left panel]) and now $U_{\rm e}/U_{\rm B}\approx43$.\\
When the SED in the flaring period is modeled considering SSC emission (red solid line in Fig. \ref{SEDnuclear} [right panel]) the electron distribution as well as the magnetic field should vary. As the X-ray spectrum is soft ($\Gamma_{\rm X}=1.8$), the modeling yielded a lower $E^{\prime}_{\rm min}=(67.57\pm19.38)$ MeV and $\alpha\simeq2.79\pm0.15$. As the \gray spectrum is shifted to higher energies, then $E^{\prime}_{\rm cut}=(6.32\pm2.32)\:{\rm GeV}$ cutoff is required. Since the emitting region radius decreases $\sim25.6$ times, in order to produce a synchrotron flux of the same order (or higher), the magnetic field should increase ($0.86\pm0.1$ G), because the synchrotron emission depends on the total number of the emitting electrons $N_{\rm e}$, $\delta$ and magnetic field $B$. In this case, the required electron energy density exceeds that of the magnetic field only 2.6 times, meaning there is no significant deviation from equipartition. The radio data are also plotted as reference values, but we note that in the flaring state the radio flux can also increase. However, the model does not predict a flux that significantly exceeds the observed radio data.\\
In the right panel of Fig. \ref{SEDnuclear}, SSC (blue dashed) and EIC (blue dot dashed) modeling of the SED is shown (blue solid line corresponds to SSC+EIC). The X-ray emission can be explained by the SSC emission produced in a blob of the size similar to that emitting in a quiescent state, but additional changes in $\alpha$, $E^{\prime}_{\rm cut}$ and $B$ are necessary to account for the new X-ray spectrum. Whereas the \gray emission is entirely due to the IC scattering of external photons in the fast and compact blob which is strongly particle-dominated with $U_{\rm e}/U_{\rm B}\approx519$ and the magnetic field $B=0.11\pm0.1$ G which does not differ much from the values obtained in the one-zone models. In the radio band, the modeling predicts a higher flux as compared with the presented radio data. As the radio data are not from the source active periods, this is not a strong argument to disfavor such modeling. When a larger value for $E^{\prime}_{\rm min}$ is used in the modeling, it does not introduce significant changes in the model parameters (especially in the energetics of the jet). Even if the data are not sufficient for the estimation of the parameters with a high significance, so the conclusions are not definite, this is an interesting modeling, as it could possibility explain the rapid \gray activities. Such blob can be naturally formed in the reconnection events that could produce compact regions of rapidly moving plasma inside the jet (''jet in jet scenario'', \citep{giannios1,giannios2}).
\subsection{The origin of emission from knots}
The black lines in Fig. \ref{knot_spectra} represents the IC/CMB radiation model calculated for $\delta=10$. The obtained parameters are presented in Table \ref{table_fit_knot}.
Similar photon indicies observed in the radio and X-ray bands allowed to well define the power-law index of electrons which varies from $2.4$ to $2.6$. The cut-off energy estimated for k4 is $E_{\rm c}^{\prime}\approx (916.3\pm251.4)$ GeV implies there is an effective particle acceleration above the TeV energies. $E^{\prime}_{\rm min}\simeq(18.20-24.01)$ MeV is estimated, which we obtain by requiring a turnover below the X-ray data in order to not overproduce the radio/optical flux but in principle lower values cannot be excluded. The IC scattering of CMB photons with $u_{\rm CMB}\simeq4.0\times10^{-11}\:{\rm erg\:cm^{-3}}$ density (when $\delta=10$) still predicts a flux lower than the observed one, so we were forced to adopt larger values of $\eta$. We found that when $B=(0.93-5.6)$ ${\rm \mu G}$ and $\eta=(83.4-410.8)$, the IC/CMB model can reproduce the observed spectra.  The maximum electron energy density is estimated to be $1.12\times10^{-10}\:{\rm erg\:cm^{-3}}$ for the k4, while for the other knots it is $>4$ times lower. This is natural, since even if the total number of particles is conserved, the low-energy cutoff moves to lower energies (because of adiabatic losses), the normalization decreases and $U_{\rm e}$ does so as well. The IC/CMB component predicts emission up to $\sim\epsilon_{\rm CMB}(E_{e}/m_{e}c^2)^2\approx27.4\:{\rm GeV}$ so that \gray emission is also expected. However, even if the predicted flux is above the Fermi LAT sensitivity ($\sim10^{-13}\:{\rm erg\:cm^{-2}s^{-1}}$), its level (a few times $10^{-13}\:{\rm erg\:cm^{-2}s^{-1}}$) would be still below the core emission in the quiet state (Fig. \ref{SEDnuclear}). Moreover, the Doppler boosting of $\delta\geq10$ requires the jet to be highly relativistic or viewed at small angles at kpc distances from the core which seems unrealistic for 3C 120, so even lower flux levels are expected.\\
The blue lines in Fig. \ref{knot_spectra} show the two-component synchrotron model fitting of knots' SEDs. The radio-to-optical data of the four knots are modeled with synchrotron emission with the following plausible parameters: $B$ between 10.1 and 115.6 $\mu G$ and an electron power-law index of $\alpha=2.20-2.48$. The plasma in the knots is in equipartition, $U_{e}=U_{B}=(4.1-531.5)\times10^{-12}$ ${\rm erg\:cm^{-3}}$, which requires more than 10 times stronger magnetic field for all the knots, as compared with the previous modeling. The synchrotron emission of the second population of electrons for the same magnetic field can explain the X-ray flux when $E_{\rm min}\simeq(2.83-8.88)\:{\rm TeV}$ and $\alpha= 2.32-2.69$. The particle energy density of this component is negligible as compared with the other one. A significant contribution from the electrons with $E_{e}\simeq10\:{\rm TeV}$ is expected, the cooling time of which $t_{\rm cool}=6\:\pi \:m_e^2\:c^{3}/\sigma_{\rm T} B^2\:E_{e}\simeq255.75$ yr. This corresponds to a travel distance of $c\times t_{\rm cool}\simeq78.4\:{\rm pc}$, which is much smaller than the size of the knots. Thus, it is required that the particle acceleration in situ over the entire volume of the knots should be extremely efficient.\\
The above obtained parameters were estimated taking into account the equipartition condition, when the system is close to internal pressure or energy density condition. But for jet dynamics and propagation more important is the jet pressure balance with the ambient medium. The results presented here and previous observations of the knots allow to put important constraints on some of the physical parameters of the jet. The jet half opening angle ($\theta_{j}$) at kpc scale can be estimated using the first resolved jet knot (k4); at a distance of 4 arcseconds from the core its radius is 0.738 arcsecond, implying $\theta_{j}\simeq10.45^{\circ}$. Having the independent information on the jet Doppler factor, the upper bound on the magneto sonic (Mach) number is $M_{\rm j}\sim1/{\rm tan(\theta_{j})\:\Gamma_{\rm bulk}}\simeq5.42/\Gamma_{\rm bulk}$. If the jet remains relativistic up to kpc scale with $\Gamma_{\rm bulk}=5.3\pm1.2$, $M_{\rm j}\simeq1.02$. For the pc jet of 3C 120, assuming an $R=1.56\times10^{16}\:{\rm cm}$ emitting region at parsec distance, $M_{\rm j}$ corresponds to $37.32$. Thus, the relativistic jet with an initial high Mach number comes into static pressure equilibrium with the interstellar medium of the parent galaxy, starting to interact with it, causing the Mach number to decrease. This is qualitatively supported by the radio/X-ray observations, which reveal that at the distance of after k4/k7 knots the jet starts to expand (e.g., Fig. 3 of \citet{hariss04}).
\subsection{Jet energetics}
The fundamental quantity is the total power (particles + magnetic field) transported by the jet flow. The total jet power can be estimated using the parameters derived from the SEDs modeling by $L_{e}=\pi c R_b^2 \Gamma^2 U_{\rm e}$ and $L_{\rm B}=\pi c R_b^2 \Gamma^2 U_{\rm B}$ \citep{cellot} for electrons and magnetic field, respectively ($\Gamma=1$ is assumed in the two-component synchrotron model). The protons with unknown contribution to the jet have not been considered in the calculations, since a number of assumptions need to be made.\\
In the quiescent state, the total power at the jet core for all models presented in Fig. \ref{SEDnuclear} is $L_{\rm jet}=L_{\rm e}+L_{\rm B}\simeq(2.35-48.0)\times 10^{44}\:{\rm erg \: s^{-1}}$ (like in \citep{tanaka3c120,sikora16,sahak3c120}). Thus, the isotropic \gray luminosity, $L_{{\rm em,}\gamma}\simeq8.2\times 10^{41}$ ${\rm erg\: s^{-1}}$, is only the small fraction of the total jet power. The jet’s total power, $L_{\rm jet}\simeq1.31\times 10^{44} \:{\rm erg \:s^{-1}}$, decreases in the active state, since it scales with the emitting region size ($L_{\rm jet}\sim R^{2}\:U$) and a smaller region is considered. However, this region is more energetic, as the particle energy density is $\sim146.9$ times higher than that in the quiet state. The SSC+EIC scenario requires a total jet luminosity of $L_{\rm jet}\simeq4.14\times10^{45}\:{\rm erg\: s^{-1}}$, which is higher than the previous values, but is well achievable for the black hole mass in 3C 120.\\
When the jet power is estimated for the knots, their largest reasonable volumes are used, so that the obtained values are the upper limits. In case of the beamed IC/CMB scenario the total jet power should be $L_{\rm jet}\simeq(1.4-131.4)\times10^{44}$ ${\rm erg\: s^{-1}}$ in order to explain the X-ray luminosity of $L_{\rm X}\simeq (1.0-4.0)\times10^{41}$ ${\rm erg\: s^{-1}}$. This jet luminosity is mostly defined by the kinetic energy of particles since the modeling reveals a moderate domination of particles over the magnetic field ($\eta>>1$).
In the two-component synchrotron model, the total jet luminosity is lower, $L_{\rm jet}\leq 6.7\times10^{44}$ ${\rm erg\: s^{-1}}$, where the contribution of the X-ray emitting component is negligible. The powers independently derived for the inner and outer regions of the jet are of the same order, suggesting that the jet does not substantially dissipate its power until its end but it becomes radiatively inefficient farther from the formation point.
\section{Summary}\label{sec6}
The main properties of the powerful jet of 3C 120 are investigated by comparing the physical state of the plasma on sub-pc and kpc scales. The main processes responsible for the broadband emission in the innermost ($\leq$ pc; Swift XRT/UVOT and Fermi LAT data) and outer ($\geq1$ kpc; Chandra data) regions are also studied.\\
On April 24, 2015, a rapid and dramatic increase of the \gray flux was observed from the inner jet of 3C 120. Within 19.0 min and 3.15 hours the flux was as high as $(7.46\pm1.56)\times10^{-6}\;{\rm photon\:cm^{-2}\:s^{-1}}$ and $(4.71\pm0.92)\times10^{-6}\;{\rm photon\:cm^{-2}\:s^{-1}}$ above 100 MeV which corresponds to an isotropic \gray luminosity of $(1.2-1.6)\times10^{46}\:{\rm erg\:s^{-1}}$. Such luminosity is unusual for radio galaxies and more typical for BL Lacs. The synchrotron/SSC mechanism gives a reasonable explanation of the multiwavelength SED in the quiescent and flaring states. The increase and rapid changes in the flaring state can be also explained assuming an additional contribution from the blob where the dominant photon fields are of external origin. The necessary jet kinetic power is $L_{\rm jet}\simeq(1.31-48.0)\times 10^{44} \:{\rm erg \:s^{-1}}$.\\
The X-ray emission from the knots has a hard photon index of $\simeq(1.6-1.8)$ with a luminosity of $L_{\rm X}\simeq (1.0-4.01)\times10^{41}\:{\rm erg\: s^{-1}}$. This X-ray emission can be explained by IC/CMB models only if $\delta>10$, otherwise the particle energy density will strongly dominate over that of the magnetic field. If the X-rays are produced from the direct synchrotron radiation of the second population of electrons, which are produced more recently than the cooler population responsible for the radio-to-optical spectrum, lower jet luminosity and no bulk relativistic motion on kpc scales is required.\\
The jet luminosities of the innermost and outer regions are comparable, suggesting that the jet does not suffer important energy losses from the regions close to the black hole to those at hundreds of kiloparsecs from it. However, at larger distances the magnetic field and the particle energy density decrease and the jet becomes radiatively inefficient.
\section*{Acknowledgments}
This work was supported by the RA MES State Committee of Science, in the frames of the research project No 15T-1C375. Also, this work was made in part by a research grant from the Armenian National Science and Education Fund (ANSEF) based in New York, USA.
\bibliographystyle{aa}
\bibliography{biblio}{}
\end{document}